\documentclass[a4paper,11pt]{article}
\usepackage{jheppub} 
\usepackage{lineno}

\title{Fine-tuning problems in type IIA string theory}

\author[1,2]{Yang Liu}
\affiliation[1]{School of Physics and Astronomy, University of Nottingham, University Park, Nottingham NG7 2RD, United Kingdom}
\affiliation[2]{Nottingham Centre of Gravity, University of Nottingham, Nottingham NG7 2RD, UK}

\emailAdd{yang.liu@nottingham.ac.uk}

\abstract{We demonstrate a unified resolution to the strong CP, hierarchy, and cosmological constant problems in type IIA flux compactifications, via 4-form fluxes and KL stabilization. We show that the strong CP problem can be effectively “solved” in type IIA orientifold constructions, particularly in the type IIA $T^6/(\mathbb{Z}_2 \times \mathbb{Z}_2)$ model. Building on this, we explore whether the remaining two fine-tuning problems can also be resolved within the same setup. To obtain a small cosmological constant, we adopt the KL scenario and find that, in order to avoid conflicts with the swampland distance conjecture and to eliminate the need for fine-tuning, the perturbative superpotential $\Delta W$ must take the form $f_0 U^3$. Additionally, we compute the gravitino mass. This allows for a resolution of the hierarchy problem without introducing fine-tuning if gravitino mass lies below 100 TeV. Taken together, these results suggest that the type IIA $T^6/(\mathbb{Z}_2 \times \mathbb{Z}_2)$ orientifold model provides a promising framework in which all three fine-tuning problems may be addressed simultaneously.}

\begin{document}
\maketitle
\flushbottom

\section{Introduction}

In fundamental physics, three of the most persistent fine-tuning problems -- the strong CP problem, the hierarchy problem, and the cosmological constant problem -- remain unresolved despite decades of theoretical effort. These issues are usually studied in isolation, and while promising ideas exist for each individually, a unified framework addressing all three within string theory is still lacking. In this paper, we propose a concrete mechanism in type IIA string compactifications that simultaneously tackles these problems. Our approach embeds the 4-form mechanism for axion potential generation into a toroidal orientifold model, leverages a Kallosh-Linde-type (KL) structure to stabilize moduli with a small superpotential correction $\Delta W$, and ensures compatibility with the Swampland Distance Conjecture (SDC).

Bousso and Polchinski \cite{bousso2000quantization} proposed that to address the cosmological constant problem, one must consider the contribution of 4-forms to the cosmological constant within the context of string theory. They argued that string theory can give rise to multiple 4-forms, and the values of these 4-forms are quantized. If the number of 4-forms and their possible quantized values are sufficiently large, it may be possible to account for the small observed value of the cosmological constant. The scalar potential takes on a schematic form as follows:
\begin{equation} \label{Vcc}
   V = \sum_i F^2_i - V_{obs},
\end{equation}
where $F_i = \epsilon^{\mu\nu\rho\sigma} F^i_{\mu\nu\rho\sigma}$ and $-V_{obs}$ denotes the remaining contributions, typically yielding a negative value. In this scenario, the cancellation is not dynamical but is instead assumed based on an anthropic argument. A challenge with this approach is that, within string vacua, it is impossible to separate the cosmological constant issue from the moduli fixing problem. One would expect the 4-forms to interact with the moduli, which complicates the situation significantly. As is well known, shortly after the general approach for fixing all moduli in type IIB string theory vacua was proposed \cite{kachru2003sitter}, internal RR and NS fluxes were introduced \cite{giddings2002hierarchies} to stabilize the complex structure moduli and the dilaton in type IIB orientifolds, with the Kähler moduli assumed to be fixed by non-perturbative effects. Since then, considerable effort has been devoted to the problem of moduli fixing, particularly in relation to internal fluxes.

Four-forms were also explored in papers by Dvali \cite{dvali2006large,dvali2005three,dvali2006vacuum,dvali2014neutrino}, where it was demonstrated that the strong CP problem and its axion solution can be elegantly described using a composite 3-form, the QCD Chern-Simons term $C^{(3)}$, with the dynamical 4-form being proportional to $F \wedge F$. In this framework, the PQ solution to the strong CP problem corresponds to the 3-form acquiring mass through its coupling to a 2-form $B_{\mu\nu}$, which is the dual of a standard axion \cite{dvali2006large,dvali2005three,dvali2006vacuum,dvali2014neutrino}. Moreover, as discussed in \cite{dvali2022strong, burgess2024uv, choi2023implications}, the authors proposed that when the number of 3-form fields coupled to the QCD axion surpasses the number of 2-form fields in the theory, it is possible to resolve the strong CP problem while simultaneously avoiding the so-called “quality problem”.

While moduli stabilization and vacuum uplift have been widely studied in the context of type IIB string theory -- most notably in the KKLT scenario \cite{kachru2003sitter} and its Large Volume Scenario (LVS) extensions \cite{balasubramanian2005systematics,conlon2005large} -- these constructions typically focus on solving either the cosmological constant problem or the hierarchy problem, and rarely address the strong CP problem. On the other hand, the 4-form mechanism developed in \cite{dvali2006large,dvali2005three,dvali2006vacuum,dvali2014neutrino, dvali2022strong, burgess2024uv, choi2023implications} offers a compelling resolution to the strong CP problem but has not been systematically embedded into a string compactification with stabilized moduli. Our work fills this gap by combining these elements into a consistent type IIA framework, thereby offering a new pathway for addressing multiple fine-tuning problems in a single string model. In this paper, we will focus on type IIA orientifolds, specifically type IIA $T^6/(\mathbb{Z}_2 \times \mathbb{Z}_2)$ orientifolds.



This work represents the first step in a broader program to construct realistic, unified models of particle physics and cosmology from string theory. In this paper, we focus on the theoretical consistency of the framework and demonstrate that all three fine-tuning problems can, in principle, be addressed within a single type IIA compactification scenario. While we do not yet embed the Standard Model or make direct experimental predictions, we provide a concrete example demonstrating the internal consistency of the setup and outline the structure required for future phenomenological developments. The paper is organized as follows: in Section 2, we review a mechanism for solving the strong CP problem. Section 3 covers the role of 4-forms in type IIA orientifolds and explains how to derive the potential energy of the QCD axion discussed in section 2. In Section 4, we review the process of obtaining (near Minkowski) de Sitter vacua in the STU model. Section 5 explains how the hierarchy problem and the cosmological constant problem are “resolved” within the STU model. Finally, in Section 6, we summarize our findings and address some remaining issues.

\section{A mechanism for solving the strong CP problem}
In \cite{dvali2022strong, burgess2024uv, choi2023implications}, the authors argued that if the number of 3-forms coupled to the QCD axion exceeds the number of 2-forms in the theory, the strong CP problem can be addressed through the following mechanism. In this section we briefly review the results of \cite{burgess2024uv}. 

Initially, the Lagrangian above the QCD scale is 
\begin{eqnarray} \label{LBAC}
\begin{split} 
\mathcal{L}_1(B,A, \mathcal{C}) \supset &  -\frac{1}{2 \cdot 3!} G_{\mu\nu\lambda} G^{\mu\nu\lambda} - \frac{1}{3!} \epsilon^{\mu\nu\lambda\rho} G_{\mu\nu\lambda} J_{\rho} -\frac{1}{4} F_{\mu\nu} F^{\mu\nu} -\frac{\theta}{2} \epsilon^{\mu\nu\lambda\rho} F_{\mu\nu} F_{\lambda\rho} \\
& -\frac{1}{4!} \eta_A \mathcal{H}^A_{\mu\nu\lambda\rho} \epsilon^{\mu\nu\lambda\rho} -\frac{1}{2 \cdot 4!} \mathcal{H}^A_{\mu\nu\lambda\rho} \mathcal{H}^{\mu\nu\lambda\rho}_A,
\end{split}
\end{eqnarray}
where $A=1,...,N$, $\mathcal{H}^A=d\mathcal{C}^A$ and $G=dB+S$ for the QCD Chern-Simons 3-form that satisfies $\Omega=dS$ where $\Omega$ is a gauge invariant quantity \cite{burgess2024uv}. $\theta$ is coupling constant of gauge field $F_{\mu\nu}$. 

Next, the Lagrangian below the QCD scale can be obtained by integrating out the field $A$:
\begin{eqnarray} \label{LCB}
\begin{split} 
\mathcal{L}_1(\mathcal{C},B) =  & -\frac{1}{2 \cdot 3!} G_{\mu\nu\lambda} G^{\mu\nu\lambda} - \frac{1}{3!} \epsilon^{\mu\nu\lambda\rho} G_{\mu\nu\lambda} J_{\rho} -\frac{\bar{\theta}}{4!} \tilde{\Lambda}_{QCD}^2 \epsilon^{\mu\nu\lambda\rho} H_{\mu\nu\lambda\rho} \\
& -\frac{1}{2 \cdot 4!} H_{\mu\nu\lambda\rho} H^{\mu\nu\lambda\rho} -\frac{1}{4!} \eta_A \mathcal{H}^A_{\mu\nu\lambda\rho} \epsilon^{\mu\nu\lambda\rho} -\frac{1}{2 \cdot 4!} \mathcal{H}^A_{\mu\nu\lambda\rho} \mathcal{H}^{\mu\nu\lambda\rho}_A +...
\end{split}
\end{eqnarray}
From \eqref{LCB}, we can obtain
\begin{equation} \label{WXYA}
    \left(\frac{\partial W}{\partial X} \right)_{Y^A} =m_a a - \bar{\theta} \tilde{\Lambda}^2_{QCD} \quad and \quad \left(\frac{\partial W}{\partial Y^A} \right)_{X} = - \eta_A, 
\end{equation}
where 
\begin{equation} \label{W}
    W = \frac{1}{2}X^2 + \frac{1}{2} Y^A Y_A + ...
\end{equation}
and $X \equiv (1/4!) \epsilon^{\mu\nu\lambda\rho} H_{\mu\nu\lambda\rho}$ and $Y^A \equiv (1/4!) \epsilon^{\mu\nu\lambda\rho} \mathcal{H}^A_{\mu\nu\lambda\rho}$.

The dual Lagrangian is then given by
\begin{equation} \label{L2a}
    \mathcal{L}_2 (a) = -\frac{1}{2} (\partial a)^2 - J^{\mu} \partial_{\mu} a - \frac{1}{2} J_{\mu} J^{\mu} - V(a),
\end{equation}
where
\begin{eqnarray} \label{Va}
\begin{split} 
V(a) & = -W(X, Y^A) + (m_a a - \bar{\theta} \tilde{\Lambda}^2_{QCD})X  - \eta_A Y^A\\
& = - \frac{1}{2} X^2 + (m_a a - \bar{\theta} \tilde{\Lambda}^2_{QCD})X + \frac{1}{2} \eta_A \eta^A.\\
\end{split}
\end{eqnarray}
The term $(m_a a - \bar{\theta} \tilde{\Lambda}^2_{QCD})X$ violates CP symmetry. From eq.\eqref{Va}, we obtain
\begin{equation} \label{VaX}
    \frac{\partial V(a)}{\partial a} =m_a X.
\end{equation}

From equation \eqref{VaX}, we observe that $X=0$ in the vacuum, which implies that 
\begin{equation} \label{Vvacuum}
    \left(\frac{\partial W}{\partial X} \right)_{Y^A} = X = m_a a - \bar{\theta} \tilde{\Lambda}^2_{QCD}=0.
\end{equation}
As a result, the term $(m_a a - \bar{\theta} \tilde{\Lambda}^2_{QCD})X$ vanishes, ensuring that the strong CP problem remains resolved \cite{burgess2024uv}. Additionally, this mechanism can also circumvent the “quality problem”\cite{burgess2024uv}. Similar mechanism can also be found in \cite{dvali2022strong,choi2023implications}.

Moreover, we can extend the above mechanism to the case of multiple axions \cite{burgess2024uv}. For the two-axion scenario, the Lagrangian below the QCD scale is
\begin{equation} \label{L2ab}
   \mathcal{L}_2 (a,b) = -\frac{1}{2} (\partial b)^2 - \frac{1}{2} (\partial a +J)^2 - V(a,b)
\end{equation}
with
\begin{equation} \label{Vab}
   V (a,b) = -W(X, Y) + (m_a a - \bar{\theta} \tilde{\Lambda}^2_{QCD})X + (\tilde{m}_a a + m_{\star} b - \eta \tilde{\Lambda}^2_{X})Y,
\end{equation}
where $X$ and $Y$ have been defined as before. For the simplest example of $W = \frac{1}{2}(X^2 + \frac{1}{2} Y^2)$, we can obtain
\begin{equation} \label{WabX}
    \left(\frac{\partial W}{\partial X} \right)_{Y^A} = m_a a - \bar{\theta} \tilde{\Lambda}^2_{QCD}
\end{equation}
and
\begin{equation} \label{WabY}
    \left(\frac{\partial W}{\partial X} \right)_{Y^A} =\tilde{m}_a a + m_{\star} b - \eta \tilde{\Lambda}^2_{X}.
\end{equation}
Differentiating \eqref{Vab} with respect to $a$ and $b$, we have
\begin{equation} \label{DVab}
    \frac{\partial V(a,b)}{\partial a} = m_a X + \tilde{m}_a Y \quad \text{and} \quad \frac{\partial V(a,b)}{\partial b} = m_{\star} Y.
\end{equation}
Therefore, all extrema of the potential satisfy $X=Y=0$. Further details about this mechanism can be found in \cite{dvali2022strong, burgess2024uv, choi2023implications}.

In \cite{dvali2022strong}, the author argued that only massive 3-forms are compatible with gravity, which necessitates the presence of an axion. Furthermore, the current understanding of quantum gravity is primarily based on its S-matrix formulation, which imposes stringent constraints on the theory's vacuum landscape. Specifically, this framework rules out the possibility of de Sitter vacua, i.e., vacua with positive energy densities. However, other cosmological spacetimes that do not asymptotically approach the Minkowski vacuum are also problematic from the S-matrix perspective \cite{dvali2022strong}. In fact, in \cite{Liu:2023vqp,Liu:2024blx}, the authors discovered that Minkowski vacua are the most stable vacua for a broad class of effective field theories. Consequently, in this paper, we begin with a curved spacetime that is very close to Minkowski spacetime. de Sitter and anti-de Sitter vacua are treated as small perturbations around the Minkowski spacetime.

Here we make a comment. While the 4-form-coupled axion mechanism resolves the strong CP problem in effective field theory, its full consistency requires embedding into quantum gravity. In Section 3.2, we achieve this in type IIA string theory via compactification on the $T^6/(\mathbb{Z}_2 \times \mathbb{Z}_2)$ orientifold. Here, RR/NS fluxes induce Minkowski 4-forms \eqref{starF042}-\eqref{starH04} that couple to the QCD Chern-Simons term \eqref{F4aaxion}, naturally generating the axion potential \eqref{Vab}. Crucially, the flux constraint \eqref{hi} enforces $X=0$ in the vacuum, dynamically realizing $\bar{\theta}$ within string theory.

\section{4-forms in type IIA orientifolds and the strong CP problem}
In this section, we will first examine 4-forms in type IIA orientifolds. Next, we will discuss how to derive the potential energy of the QCD axion from type IIA orientifolds. 

\subsection{Review of 4-forms in type IIA orientifolds}
We review the appearance of $4D$ 4-forms in type IIA orientifold compactifications. The compactification of ten-dimensional massive type IIA string theory on a Calabi-Yau threefold with background fluxes has been extensively studied in \cite{grimm2005effective, louis2002type, villadoro2005N, dewolfe2005type, camara2005fluxes}. In \cite{bielleman2015minkowski}, the authors performed the same compactification, while carefully tracking all the Minkowski 4-forms that emerge upon dimensionally reducing the $10D$ RR and NSNS fields. This results in a new formulation of the scalar potential in terms of Minkowski 4-forms \cite{bielleman2015minkowski}.

\subsubsection{4-forms, RR and NS fluxes in IIA orientifolds}

We are interested in the presence of Minkowski 3-form fields in the flux-induced scalar potential. In addition to the universal RR 3-form $C_3$, one can also obtain 3-forms by dimensionally reducing higher RR and NSNS fields, such as $C_5$ , $C_7$ , $C_9$ and $H_7$, while considering three of the indices in Minkowski space. We will work in the democratic formulation \cite{bergshoeff2001new}, where all the $p$-form fields $C_p$ with $p = 1, 3, 5, 7$ are included. As a result, we will need to impose the Hodge duality relations
\begin{equation} \label{Hodgeduality}
  G_6 = - \star_{10} G_4, \quad G_8 =  \star_{10} G_2, \quad G_{10} =  -\star_{10} G_0,
\end{equation}
at the level of the equations of motion, we impose conditions to avoid overcounting the physical degrees of freedom. As a result, we obtain $2 h^{(1,1)}_{-}+2$ Minkowski 4-forms: $F^0_4$, $F^i_4$, $F^a_4$ and $F^m_4$. Specifically, there are $h^{(1,1)}_{-}$ $F^i_4$ fluxes, $h^{(1,1)}_{-}$ $F^a_4$ fluxes, one $F^0_4$ flux and one $F^m_4$ flux. The details are not discussed in this paper, but further information can be found in \cite{bielleman2015minkowski}. Additionally, the fields $B_2$ and $C_3$ can be expanded as follows:
\begin{equation} \label{B2C3basis}
  B_2 = \sum_i b_i \omega_i, \quad C_3= \sum_I c^I_3 \alpha_I.
\end{equation}
Here $b_i$ and $c^I_3$ are $4D$ scalars that represent the axionic components of the complex supergravity fields $T$, $S$, $U$, as given by the following:
\begin{equation} \label{ImTi}
 Im T_i= - \int B_2 \wedge \tilde{\omega}^i = -b^i; \quad i= 1,..., h^{(1,1)}_{-}
\end{equation}
\begin{equation} \label{ImUi}
 Im U_i= \int C_3 \wedge \beta^i = c^i_3; \quad i= 1,..., h^{3}_{+}
\end{equation}
\begin{equation} \label{ImS}
 Im S= -\int C_3 \wedge \beta^0 = -c^0_3,
\end{equation}
where $\tilde{\omega}^i$, $\beta^i$ and $\beta^0$ are the elements of the cohomology basis \cite{bielleman2015minkowski}. By decomposing each field strength into its Minkowski and internal components, the duality relations provided in \eqref{Hodgeduality} imply:
\begin{equation} \label{starF04}
  \star_4 F^0_4 = \frac{1}{k} \left(e_0 + e_ib^i + \frac{1}{2} k_{ijk} q^i b^j b^k - \frac{m}{3!} k_{ijk} b^i b^j b^k -h_0 c^0_3 - h^i c^i_3\right),
\end{equation}
\begin{equation} \label{starFi4}
  \star_4 F^i_4 = \frac{g^{ij}}{4k} \left(e_i + k_{ijk} b^j q^k -\frac{m}{2} k_{ijk} b^j b^k  \right),
\end{equation}
\begin{equation} \label{starFa4}
  \star_4 F^a_4 = 4kg^{ab} (q_b -m b_b),
\end{equation}
\begin{equation} \label{starFm4}
  \star_4 F^m_4 = -m.
\end{equation}
Here $g_{ij} = \frac{1}{4k} \int \omega_i \wedge \star \omega_j $ represents the metric in the Kähler moduli space, where $k$ is the volume and $k_{ijk}$ is the topological triple intersection number \cite{bielleman2015minkowski}.

The type IIA ten dimensional supergravity action can be expressed as the sum of three terms \cite{bielleman2015minkowski},
\begin{equation} \label{SIIA}
  S_{IIA}=S_{RR} + S_{NS} + S_{loc},
\end{equation}
where the $S_{RR}$ and $S_{NS}$ are the RR and NSNS actions, respectively, while $S_{loc}$ refers to the contribution from localized sources such as $D6$-branes and $O6$-planes. 

By integrating over the internal dimensions, we obtain the following effective scalar potential in $4D$ \cite{bielleman2015minkowski},
\begin{eqnarray} \label{VRR}
\begin{split} 
V_{RR} =  & -\frac{1}{2} [-k F^0_4 \wedge \star F^0_4 +2F^0_4 \rho_0 - 4k g_{ij} \star F^i_4 \wedge F^j_4 + 2 F^i_4 \rho_i   \\
& -\frac{1}{4k} g_{ab} F^a_4 \wedge \star F^b_4 + 2 F^a_4 \rho_a + k F^m_4 \wedge \star F^m_4 ],
\end{split}
\end{eqnarray}
where all 4-four fluxes are provided in \eqref{starF04}-\eqref{starFm4}, and the Chern-Simons couplings of the 4-forms are given by
\begin{equation} \label{rho0}
  \rho_0 = e_0 + b^i e_i + \frac{1}{2} k_{ijk} q_i b^j b^k - \frac{m}{6} k_{ijk} b_i b^j b^k - h_0 c^0_3 - h_i c^i_3,
\end{equation}
\begin{equation} \label{rhoi}
  \rho_i = e_i + k_{ijk} b^j q^k - \frac{m}{2} k_{ijk} b^j b^k,
\end{equation}
\begin{equation} \label{rhoa}
  \rho_a = q_a - m b_a,
\end{equation}
\begin{equation} \label{rhom}
  \rho_m =- m.
\end{equation}
The potential $V_{NS}$ is
\begin{equation} \label{VNS}
  V_{NS}=\frac{1}{2} e^{-2\phi} c_{IJ} H^I_4 H^J_4,
\end{equation}
where $c_{IJ}$ is the metric on the complex structure moduli space and the NS internal flux satisfies $\star H^I_4 =h^I$ \cite{bielleman2015minkowski}. 

The potential $V_{loc}$, arising from the localized sources, is given by:
\begin{equation} \label{Vloc1}
  V_{loc}=\sum_a \int_{\Sigma} T_a \sqrt{-g} e^{-\phi},
\end{equation}
where $T_a$ is the tension of brane and $\Sigma$ the worldvolume. 

\subsubsection{4-forms and geometric fluxes in type IIA $T^6/(\mathbb{Z}_2 \times \mathbb{Z}_2)$ orientifolds}
It is known that, besides the standard RR and NS fluxes, there may also be other less-explored NS fluxes. These include the geometric fluxes in toroidal models, which arise in the context of Scherk-Schwarz reductions \cite{bielleman2015minkowski}. 

We are interested in how the presence of geometric fluxes affects the 4-forms described in equations \eqref{starF04}-\eqref{starFm4}. Geometric fluxes can be defined on a factorized 6-torus $T^6$, where O6-planes wrap 3-cycles. Additionally, we assume a $\mathbb{Z}_2 \times \mathbb{Z}_2$ orbifold twist, which results in only diagonal moduli surviving the projection. In this case, we are left with 3 Kähler moduli and 4 complex structure moduli (including the complex dilaton). This setup involves 12 geometric fluxes $\omega^M_{NK}$, which can be conveniently organized in a 3-vector $a_i$ and a $3 \times 3$-matrix $b_{ij}$ \cite{bielleman2015minkowski}. 

As a result, the 4-forms can be altered as follows \cite{bielleman2015minkowski}
\begin{equation} \label{starF042}
  \star F^0_4 = \frac{1}{k} \left(e_0 + e_ib^i + \frac{1}{2} k_{ijk} q^i b^j b^k - \frac{m}{3!} k_{ijk} b^i b^j b^k -h_0 c^0_3 - h^i c^i_3 + b^i b_{ij} c^j_3 - b^i a_i c^0_3 \right),
\end{equation}
\begin{equation} \label{starFi42}
  \star F^i_4 = \frac{g^{ij}}{4k} \left(e_i + k_{ijk} b^j q^k -\frac{m}{2} k_{ijk} b^j b^k  b_{jk} c^k_3 - a_j c^0_3 \right),
\end{equation}
\begin{equation} \label{starHi4}
  \star H^i_4 = h^i - b^{ij}b_j,
\end{equation}
\begin{equation} \label{starH04}
  \star H^0_4 = h^0+b^i a_i.
\end{equation}
Here, $k_{ijk}$ are the intersection numbers, which equal one if all the indices are distinct and zero otherwise. 

For $T^6/(\mathbb{Z}_2 \times \mathbb{Z}_2)$ orientifold, the potential $V_{D6/O6}$ coming from localized sources can be conveniently rewritten by exploiting the requirement that the brane setup preserves $N = 1$ supersymmetry, namely,
\begin{equation} \label{VD6O6}
  V_{D6/O6} = e^K u_1 u_2 u_3 \sum_a T_a (n^a_1 n^a_2 n^a_3 s - n^a_1 m^a_2 m^a_3 t_1 - m^a_1 n^a_2 m^a_3 t_2 - m^a_1 m^a_2 n^a_3 t_3).
\end{equation}
Here, $n^i_a \ (m^i_a)$ denotes the wrapping numbers along the $x^i \ (y^i)$ directions of the $i$-th two-torus. In this context, $s$, $t_i$ and $u_i$ are the real parts of the moduli $S$, $T_i$ and $U_i$, respectively \cite{villadoro2005N}.

\subsection{Potential energy of QCD axion}
Let us begin by examining the 4-forms in type IIA orientifolds, starting with the term 
\begin{equation} \label{F4aaxion}
-\frac{1}{4k} g_{ab} F^a_4 \wedge \star F^b_4 + 2 F^a_4 \rho_a.
\end{equation}
By comparing \eqref{starFa4}, \eqref{VRR} and \eqref{rhoa}, we can find that the term \eqref{F4aaxion} corresponds to the potential energy of the QCD axion, assuming $\frac{1}{4k} g_{ab}=1$, which leads to \eqref{Va}. This suggests that, in principle, 4-form fluxes arising in type IIA orientifold compactifications to four dimensions can naturally give rise to a potential capable of addressing the strong CP problem.

Next, let us consider the case of 4-forms in toroidal type IIA orientifolds. By comparing \eqref{starF04}-\eqref{starFm4} with  \eqref{starF042}-\eqref{starH04}, we can identify the following correspondences: 
\begin{equation} \label{4formcorresponding}
\star_4 F^0_4 \leftrightarrow  \star F^0_4, \quad \star_4 F^i_4 \leftrightarrow \star F^i_4, \quad \star_4 F^a_4 \leftrightarrow \star H^i_4, \quad \star_4 F^m_4 \leftrightarrow \star H^0_4,
\end{equation}
where the symbol “$\leftrightarrow$” denotes that the two sides are equivalent and interchangeable.

The potential $V_{RR}$ associated with the 4-form $H^i_4$, denoted as $V_{RR}(H^i_4)$, is given by:
\begin{eqnarray} \label{VRRHi4}
\begin{split} 
V_{RR}(H^i_4) \sim  & -\frac{1}{2} (- H^i_4 \wedge \star H^i_4 + 2H^i_4 \rho_i) \\
 = & -\frac{1}{2} ( -H^i_4 \wedge \star H^i_4 ) + (b_{ij} b^j - h_i) H^i_4.
\end{split}
\end{eqnarray}
By comparing \eqref{VRRHi4} with \eqref{Va}, we can identify the following correspondences:
\begin{equation} \label{axioncorresponding}
H^i_4 \leftrightarrow  X, \quad b_{ij} \leftrightarrow m_a, \quad b^i \leftrightarrow a, \quad h_i \leftrightarrow \bar{\theta} \tilde{\Lambda}^2_{QCD}.
\end{equation}
Since the mechanism discussed in Section 2 provides a resolution to the strong CP problem, it follows that:
\begin{equation} \label{hi}
b_{ij} b^j = h_i.
\end{equation}
This flux constraint \eqref{hi} is sufficient to solve the strong CP problem in string theory, with its physical realization enforced by tadpole conditions \eqref{tcc1}-\eqref{tcc4}. To conclude, by considering eqs.\eqref{Va}, \eqref{starFa4}, \eqref{starFm4}, \eqref{starHi4}, \eqref{starH04}, \eqref{4formcorresponding} and \eqref{axioncorresponding}, we obtain the following correspondences:
\begin{equation} \label{qbthetalambdahi}
q_b \leftrightarrow \bar{\theta} \tilde{\Lambda}^2_{QCD} \leftrightarrow h^i,
\end{equation}
\begin{equation} \label{mabijh0biai}
-m_a \leftrightarrow -b_{ij} \leftrightarrow h^0 + a^i b_i,
\end{equation}
\begin{equation} \label{abbbj}
a \leftrightarrow b_b \leftrightarrow b_j.
\end{equation}
Combining eqs.\eqref{qbthetalambdahi}-\eqref{abbbj}, we have the correspondence:
\begin{equation} \label{hibijbjhih0biaibj}
h^i - b^{ij} b_j \leftrightarrow h^i + (h^0 + a^i b_i) b_j,
\end{equation}
namely,
\begin{equation} \label{aibjbij}
h_0 b_j + a_i b^i b_j = - b^i b_{ij}=-h_j,
\end{equation}
where the last equation is from \eqref{hi}.

In particular, when three axions are present -- as is the case in the STU model -- the system admits the following structure:
\begin{equation} \label{bij33nonzero}
b_{11} =m_1, \quad b_{21}=m_2, \quad b_{22}=m_3, \quad b_{31}=m_4, \quad b_{32}=m_5, \quad b_{33}=m_6, 
\end{equation}
and
\begin{equation} \label{bij33zero}
b_{12} =b_{13}=b_{32}=0.
\end{equation}
From \eqref{hi}, we obtain
\begin{equation} \label{h1}
h_1=m_1 a,
\end{equation}
\begin{equation} \label{h2}
h_2=m_2 a + m_3 b,
\end{equation}
and 
\begin{equation} \label{h3}
h_3=m_3 a+m_4 b+m_5 c.
\end{equation}

\section{STU model}

To simultaneously embed the above strong CP mechanism into a specific moduli-stabilization framework and interface with cosmological constant tuning, we adopt the minimal computable framework: the STU model. With its explicit Kähler potential \eqref{Ktot1} and KL-type superpotential structure \eqref{Wtot} and \eqref{Wi}, this model provides an ideal laboratory for unifying solutions to all three fine-tuning problems while enabling controlled anti-de Sitter downshifts and de Sitter uplifts \eqref{VbarD6}. The aim of this section is to present this model and explain its components \cite{cribiori2019uplifting}.

In the notation adopted from \cite{cribiori2019uplifting,dibitetto2011charting,danielsson2014alternative}, $S$ represents the axio-dilaton, $T$ is a complex structure modulus, and $U$ is the volume (Kähler) modulus. We propose utilizing ten-dimensional supergravity compactified on a calibrated manifold, such as a Calabi-Yau manifold or a more general SU(3)-structure manifold, so that the standard four-dimensional $\mathcal{N} = 1$ supergravity is linearly realized. This framework will be enhanced by pseudo-calibrated $\overline{D6}$-branes to enable a KKLT-like uplift \cite{cribiori2019uplifting}. 

To make our example more concrete, we examine a $T^6/(\mathbb{Z}_2 \times \mathbb{Z}_2)$ orbifold compactification of type IIA string theory, along with the ten-dimensional metric \cite{cribiori2019uplifting}
\begin{equation} \label{10dmetric}
  ds^2_{10} = \tau^{-2} ds^2_4 + \rho( \sigma^{-3} G_{ab} dy^a dy^b + \sigma^3 G_{ij} dy^i dy^j). 
\end{equation}
Here, the universal moduli $\rho$, $\tau$ and $\sigma$ are identified as
\begin{equation} \label{rho}
  \rho = Re \ U  = (vol_6)^{1/3},
\end{equation}
\begin{equation} \label{tau}
  \tau = (Re \ S)^{1/4} (Re \ T)^{3/4} = e^{-\phi} \sqrt{vol_6}, 
\end{equation}
\begin{equation} \label{sigma}
 \sigma = (Re \ S)^{-1/6} (Re \ T)^{1/6},
\end{equation}
where the moduli are $T_i =t_i + i\theta_i$ and $Re$ represents the real part of moduli $t_i$, while $G_{ab}$ and $G_{ij}$ correspond to the two independent three-cycles. To this construction, we include $N^{\parallel}_{\overline{D6}}$ and $N^{\perp}_{\overline{D6}}$ anti-D6-branes wrapping three-cycles. The first set of branes extends fully along a single cycle, while the second set corresponds to branes wrapping along both cycles in all possible combinations.

After compactifying to four dimensions, the total scalar potential is the sum of two components \cite{cribiori2019uplifting}
\begin{equation} \label{Vtot}
 V_{tot} = V_{\mathcal{N}=1} + V_{\overline{D6}},
\end{equation}
where
\begin{equation} \label{VN1}
  V_{\mathcal{N}=1} = e^{K} (g^{i \bar{j}} D_i W D_{\bar{j}} W - 3|W|^2), \quad i=\{S,T,U\},
\end{equation}
is the standard $\mathcal{N} = 1$ supergravity scalar potential and 
\begin{equation} \label{VbarD6}
 V_{\overline{D6}} = \frac{\mu^4_1}{(Re \ T)^3} + \frac{\mu^4_2}{(Re \ T)^2 (Re \ S)}
\end{equation}
is the contribution of the anti-D6-branes. The quantities $\mu^4_1 = 2 e^{\mathcal{A}_1} N^{\parallel}_{\overline{D6}}$ and $\mu^4_2 = 2 e^{\mathcal{A}_2} N^{\perp}_{\overline{D6}}$ represent anti-D6-branes wrapped on two different types of three-cycles, which are located in potentially warped regions characterized by warp factors $e^{\mathcal{A}_1}$ and $e^{\mathcal{A}_2}$, respectively \cite{cribiori2019uplifting}. 

The KL generalization of the STU model discussed in \cite{cribiori2019uplifting} can be expressed, using slightly different notation, as:
\begin{equation} \label{Wtot}
 W= W_0 + \sum_{i=1}^3 W_i (T_i) + \Delta W +\mu^2 X,
\end{equation}
\begin{equation} \label{Ktot1}
 K= -\ln(T_1 + \bar{T}_1) - 3 \ln(T_2 + \bar{T}_2) - \ln \left( (T_3 + \bar{T}_3)^3 - \frac{X \bar{X}}{(T_1 + \bar{T}_1) + g(T_2 + \bar{T}_2)}\right).
\end{equation}
Here $T_1$ represents the field $S$, $T_2$ stands for the field $T$, $T_3$ corresponds to the field $U$, $W_0=f_6$ is a constant flux\footnote{The flux $W_0$ is not required to be constant; it may exhibit dependence on the moduli. We will discuss this in Section 5.} and $X$ is a nilpotent field. The term $\frac{X \bar{X}}{(T_1 + \bar{T}_1) + g(T_2 + \bar{T}_2)}$ inside the logarithm represents the uplifting contribution from the $\overline{D6}$ brane, where $g$ is a constant. Since $X$ is a nilpotent field with the property $X^2 = 0$, the Kähler potential can be equivalently written as  
\begin{equation} \label{Ktot2}
 K= -\ln(T_1 + \bar{T}_1) - 3 \ln(T_2 + \bar{T}_2) - 3\ln (T_3 + \bar{T}_3) - \frac{X \bar{X}}{(T_3 + \bar{T}_3)^3 \left((T_1 + \bar{T}_1) + g(T_2 + \bar{T}_2)\right)}.
\end{equation}
The final term in this expression accounts for the uplifting contribution to the potential, which can be expressed as in \eqref{VbarD6}. In the final calculation, we will set $X=0$.


A specific set of superpotentials for the STU model, motivated by string theory, is described by the nonperturbative KL superpotentials \cite{kallosh2020mass}
\begin{equation} \label{Wi}
W_i (T_i) = A_i e^{-a_i T_i} - B_i e^{-b_i T_i}.
\end{equation}
In order to find a supersymmetric Minkowski vacuum at $T_i = T^0_i$, it is necessary to have
\begin{equation} \label{WT0i}
W (T^0_i) =W_0 + \sum^{n}_{i=1} W_i (T^0_i)
\end{equation}
and
\begin{equation} \label{partialWi}
\partial_{T_i} W_i = 0
\end{equation}
at $T_i=T^0_i$. Without loss of generality, we can restrict our analysis to the supersymmetric Minkowski vacuum where $\theta_i=0$. Thus, we have
\begin{equation} \label{titi0}
t_i =t^0_i = \frac{1}{a_i -b_i} \ln{\frac{a_i A_i}{b_i B_i}}.
\end{equation}
If the small correction $\Delta W$ is considered, the minimum of the potential moves to the AdS state, with $V_{AdS}$ proportional to $-(\Delta W )^2$, as demonstrated in 
\begin{equation} \label{VAdS}
V_{AdS} \approx -3 (\Delta W)^2 \prod^n_{j=1} (2t^0_j)^{-N_j}.
\end{equation}
It is important to note that this general result is independent of the choice of superpotentials $W_i(T_i)$ \cite{kallosh2020mass}. 

By considering the small contribution from a nilpotent superfield $X$, the minimum can be readily uplifted to dS. The uplifted vacuum corresponds to a stable dS state for sufficiently small values of $\Delta W$ and $\mu$. The gravitino mass after uplifting to a (nearly Minkowski) dS vacuum state is given by \cite{kallosh2020mass}
\begin{equation} \label{m232}
m^2_{3/2} = e^K (\Delta W)^2 = (\Delta W)^2 \prod^n_{j=1} (2t^0_j)^{-N_j} = \frac{|V_{AdS}|}{3}.
\end{equation}

\section{The cosmological constant problem and hierarchy problem}
In \cite{kallosh2020mass} the authors suggested a procedure which allows mass production of dS vacua in the type IIA string theory. This process consists of three steps. 

The first step is to obtain supersymmetric Minkowski vacua, specifically:
\begin{equation} \label{moduliMin}
W(z^a) = 0, \quad D_a W =\partial_a W =0, \quad \text{at} \quad z_a=z^0_a.
\end{equation}
We also require that there are no flat directions. These vacua are stable and remain so even under small perturbations of the original theory \cite{kallosh2020mass}. The second step involves shifting to a supersymmetric AdS vacuum. By modifying the superpotential with the addition of a small term, $\Delta W$, the original Minkowski vacuum is transformed into a supersymmetric AdS vacuum, with $V_{AdS} = -3 e^K (\Delta W)^2$ and $m^2_{3/2} = e^K (\Delta W)^2$. Finally, the third step involves uplifting to dS, achieved by using the anti-D6-brane.



One might question why we begin with a supersymmetric Minkowski vacuum and then transition to AdS, rather than starting directly with AdS as in the standard KKLT construction. Additionally, why do we opt to downshift from the Minkowski vacuum before uplifting to dS, when it is possible to perform the uplift directly from the Minkowski vacuum \cite{kallosh2020mass}?

The reason we choose to start with a Minkowski vacuum rather than AdS is that the discrete supersymmetric Minkowski state is always a minimum and can be strongly stabilized. This property helps maintain the stability of the AdS and dS vacua that originate from the supersymmetric Minkowski state. The downshift is necessary because, without it, the vacuum with the observable cosmological constant, uplifted from the Minkowski state with $V = 0$ to the dS state with $V_{dS} = 10^{-120}$, would have $F^2 \sim 10^{-120}$, resulting in SUSY breaking being too small. A controllable downshift separates SUSY breaking from the smallness of the cosmological constant \cite{kallosh2020mass}.

Let us begin by addressing how the hierarchy problem can be explained without fine-tuning $W$ and $e^{K/2}$. As discussed in \cite{Conlon:2008zz,linde2012supersymmetry}, all viable models of low-energy supersymmetry necessitate a light gravitino, with a mass below 100TeV. In general, the gravitino mass is given by \cite{Conlon:2008zz,linde2012supersymmetry}
\begin{equation} \label{m321}
m_{3/2} = e^{K/2}  W \leq 100\text{TeV} \leq 10^{-13}M_{pl},
\end{equation}
where $M_{pl}$ denotes the Planck mass scale\footnote{While the natural scale for $m_{3/2}$ is $M_{pl}$, successful moduli stabilization requires $m_{3/2}\ll M_{pl}$.}. There are two principal approaches to achieving this suppression \cite{Conlon:2008zz,linde2012supersymmetry}:\\
1. $|W| \ll 1$ with $e^{K/2} \sim O(1)$;\\
2. $e^{K/2} \ll 1$ with $|W| \sim O(1)$.\\
Thus, to obtain a sufficiently small gravitino mass, one must finely tune either $W$ or $e^{K/2}$ to be very close to zero. This requirement is generally considered unnatural.

However, from eq.\eqref{m232}, we have
\begin{equation} \label{m322}
m_{3/2} = e^{K/2} (\Delta W) \leq 100\text{TeV} \approx 10^{-13}M_{pl}.
\end{equation}
In this scenario, the smallness of $\Delta W$ arises naturally, allowing for a light gravitino mass without the need to fine-tune $W$ and $e^{K/2}$. Therefore, by appropriately selecting $\Delta W$, supersymmetry provides a natural solution to the hierarchy problem.



The full superpotential is given by \cite{camara2005fluxes}
\begin{eqnarray} \label{Wf}
\begin{split} 
W =  & e_0 + ih_0 S + \sum^3_{i=1} [(ie_i -a_i S -b_{ii} T_i -\sum_{j\neq i} T_j)U_i- ih_i T_i]  \\
& -q_1 U_2 U_3 - q_2 U_1 U_3 - q_3 U_1 U_2 + im U_1 U_2 U_3,
\end{split}
\end{eqnarray}
where the nonperturbative contribution $W_{np}$ has been neglected.

In the context of the STU model, the moduli satisfy $T_1=T_2=T_3=T$ and $U_1=U_2=U_3=U$. In a more general setting, the superpotential can take the form $W$ \cite{cribiori2019uplifting}
\begin{equation} \label{Wgf}
W = f_6 + (h_T + r_T U)T + (h_S + r_S U)S + f_4 U + f_2 U^2 + f_0 U^3  + W_{np},
\end{equation}
where the coefficients $f_p$ $(p=0,2,4,6)$ originate from RR-fluxes, $h_{S/T}$ arise from the integration of NSNS-flux over the corresponding 3-cycles, and $r_{S/T}$ are sourced from curvature corrections of the internal manifold. In the KL scenario, the nonperturbative term $W_{np}$ is given by \eqref{Wi}. In the following analysis, we will make use of both the notations introduced in \eqref{Wf} and \eqref{Wgf}.

In general, we consider stacks of $N_a$ intersecting D6-branes wrapping the factorizable 3-cycle
\begin{equation} \label{Pia}
\Pi_a = (n^1_a, m^1_a) \otimes (n^2_a, m^2_a) \otimes (n^3_a, m^3_a),
\end{equation}
along with their corresponding orientifold images, which wrap the cycles $\otimes_i (n^i_a, -m^i_a)$. Here $n^i_a \ (m^i_a)$ denotes the wrapping numbers along the $x^i \ (y^i)$ directions of the $i$-th two-torus, respectively. 

In the case of $\mathbb{Z}_2 \times \mathbb{Z}_2$ IIA orientifold, the RR tadpole cancellation conditions in the presence of fluxes take the following form 
\begin{equation} \label{tcc1}
\sum_a N_a n^1_a n^2_a n^3_a + \frac{1}{2} (h_0 m + a_1 q_1 + a_2 q_2 + a_3 q_3) =16,
\end{equation}
\begin{equation} \label{tcc2}
\sum_a N_a n^1_a m^2_a m^3_a + \frac{1}{2} (m h_1 - q_1 b_{11} - q_2 b_{21} - q_3 b_{31}) =-16,
\end{equation}
\begin{equation} \label{tcc3}
\sum_a N_a m^1_a n^2_a m^3_a + \frac{1}{2} (m h_2 - q_1 b_{12} - q_2 b_{22} - q_3 b_{32}) =-16,
\end{equation}
\begin{equation} \label{tcc4}
\sum_a N_a m^1_a m^2_a n^3_a + \frac{1}{2} (m h_3 - q_1 b_{13} - q_2 b_{23} - q_3 b_{33}) =-16.
\end{equation}
The value $(-16)$ in the last three conditions corresponds to the RR tadpole contribution from the remaining three orientifold planes present in the $\mathbb{Z}_2 \times \mathbb{Z}_2$ set-up. If we normalize the brane tension of type $a$ to unity, i.e., $\tau_a=1$, then the total tension becomes $T_a=N_a \tau_a = N_a$ \cite{camara2005fluxes,ibanez2012string}. With this convention, the conditions in \eqref{tcc1}-\eqref{tcc4} can give $V_{D6/O6}$ in \eqref{VD6O6}.

For the STU model, $u_1=u_2=u_3$, then from \eqref{Ktot2}, we can obtain
\begin{equation} \label{eK}
e^K= \frac{1}{128 s t^3 u^3}.
\end{equation}
Consider the general uplift term \eqref{VbarD6}. From $\frac{\partial}{\partial s} V_{up}=0$, we find
\begin{equation} \label{Vuppars}
\frac{\mu^4_2}{s^2 t^2}=0,
\end{equation}
which implies $\mu^4_2=0$. Next, from the condition $\frac{\partial}{\partial t} V_{up}=0$, we obtain
\begin{equation} \label{Vuppart}
-\frac{3\mu^4_1}{t^4}-\frac{2\mu^4_2}{st^3}=0.
\end{equation}
Given that $\mu^4_2=0$, it follows that $\mu^4_1=0$. Therefore, the uplift potential vanishes. In principle, this suggests that a downshift needs to be introduced. 

\subsection{The cosmological constant problem}
The KL downshift mechanism requires a perturbative superpotential correction $\Delta W$ to destabilize the supersymmetric Minkowski vacuum into AdS vacuum \eqref{moduliMin}-\eqref{m322}. We will explore various configurations of $\Delta W$ to determine which one offers a viable solution to the cosmological constant problem.

\subsubsection{$\Delta W$ is a constant}
At tree level, we take
\begin{equation} \label{W0tree}
W_0= f_6
\end{equation}
following the notation in \eqref{Wf}, and impose
\begin{equation} \label{fluxtree}
h_0 = e_i = a_i = b_{ij} = h_i = q_i = m =0
\end{equation}
as per the notation of \eqref{Wgf}. With these assumptions, and using \eqref{tcc1}-\eqref{tcc4}, we obtain the $V_{D6/O6}$ contribution from \eqref{VD6O6}:
\begin{equation} \label{VD6O6STU}
V^{(1)}_{D6/O6}= \frac{1}{8t^3} + \frac{3}{8st^2}.
\end{equation}
This result is consistent with \eqref{VbarD6}, and therefore, we can adopt \eqref{VD6O6STU} as the uplifting term:
\begin{equation} \label{Vup1}
V^{(1)}_{up}= \frac{1}{8t^3} + \frac{3}{8st^2}.
\end{equation}
In other words, this corresponds to $\mu^4_1=\frac{1}{8}$ and $\mu^4_2=\frac{3}{8}$. We then obtain the minimum of the uplifting potential. From $\frac{\partial}{\partial s} V^{(1)}_{up}=0$, we have
\begin{equation} \label{Vuppars}
\frac{\partial}{\partial s} V^{(1)}_{up} = -\frac{3}{8 s^2 t^2}=0,
\end{equation}
which implies either $s \rightarrow \infty$ or $t \rightarrow \infty$. We lose control of the EFT. In fact, such limits violate the distance conjecture \cite{palti2019swampland,van2022lectures,agmon2022lectures}. Furthermore, from the condition $\frac{\partial}{\partial t} V^{(1)}_{up}=0$, we find
\begin{equation} \label{Vuppart}
\frac{\partial}{\partial t} V^{(1)}_{up} = -\frac{1}{t^4}-\frac{2}{st^3} =0,
\end{equation}
leading to the relation $s=-2t$, which is incompatible with the physical assumption $t_i>0$. Therefore, the uplifting term alone cannot be considered sufficient. There are two main reasons: first, the SUSY breaking scale is too small; and second, the minimum of the uplifting potential leads to theoretical inconsistency.

If we introduce a $V_{AdS}$, then, based on \eqref{m232} and \eqref{eK}, we obtain
\begin{equation} \label{VAdS1}
V^{(1)}_{AdS} = -3 e^K |\Delta W^{(1)}|^2= -\frac{3}{128 st^3 u^3} |\Delta W^{(1)}|^2,
\end{equation}
where $|\Delta W^{(1)}|$ represents a small constant flux.
The total potential $V_{dS}$ is given by
\begin{equation} \label{VdS1}
V^{(1)}_{dS}= V^{(1)}_{AdS} + V^{(1)}_{up} = -\frac{3}{128 st^3 u^3} |\Delta W|^2 + \frac{1}{8t^3} + \frac{3}{8st^2},
\end{equation}
Thus, by setting $\frac{\partial}{\partial s} V^{(1)}_{dS}=0$, we find
\begin{equation} \label{VdS1s}
|\Delta W|^2 =16 tu^3.
\end{equation}
Similarly, from $\frac{\partial}{\partial t} V^{(1)}_{dS}=0$, we get
\begin{equation} \label{VdS1t}
9|\Delta W|^2-48su^3-96tu^3=0.
\end{equation}
Next, from $\frac{\partial}{\partial u} V^{(1)}_{dS}=0$, we have
\begin{equation} \label{VdS1u}
\frac{9}{128 st^3 u^4} \rightarrow 0.
\end{equation}
By combining \eqref{VdS1s}-\eqref{VdS1u}, we obtain the relations $s=t$ and $|\Delta W|^2=16tu^3$. Consequently, we find that $s$ and $t$ approach infinity, while $u$ tends to zero. This leads to the emergence of an infinite tower of states, which contradicts both the distance conjecture and experimental observations \cite{palti2019swampland,van2022lectures,agmon2022lectures}.

Additionally, from \eqref{VdS1s}-\eqref{VdS1u}, the total potential is given by $V^{(1)}_{dS1}=\frac{1}{8s^3}$. To achieve a small cosmological constant, $s$ must tend to infinity, which also violates the swampland distance conjecture.  

We can extend this result to the general uplift term in \eqref{VbarD6}. A similar calculation shows that $\mu^4_1=\mu^4_2=0$ if we aim to avoid violating the swampland distance conjecture. However, this contradicts the expression in \eqref{Vup1}. Thus, we cannot consider a constant $\Delta W$.

Similarly, we find that $\Delta W=f_4 U$ yields analogous results. We will not explore further details of this form of $\Delta W$. Since $u$ is very small, we will henceforth focus on $\Delta W$ that depends solely on $u$.

\subsubsection{$\Delta W= f_4 U + f_2 U^2$}
Consider the following form for $\Delta W$:
\begin{equation} \label{DeltaW2}
\Delta W^{(2)}= f_4 U + f_2 U^2,
\end{equation}
which leads to
\begin{equation} \label{DeltaW22}
|\Delta W^{(2)}|^2= |f_4|^2 u^2 + |f_2|^2 u^4 + 2Re(f_2 f^{\star}_4)u^3
\end{equation}
and 
\begin{equation} \label{VAdS2}
V^{(2)}_{AdS}= -\frac{3}{128st^3u}|f_4|^2 -\frac{3}{128st^3} |f_2|^2 u -\frac{6}{128st^3} Re(f_2 f^{\star}_4).
\end{equation}
Since solving the tadpole conditions \eqref{tcc1}-\eqref{tcc4} is not straightforward, we instead focus on the general uplift term \eqref{VbarD6}. The total potential is then given by 
\begin{equation} \label{VdS2}
V^{(2)}_{dS}= -\frac{3}{128st^3u}|f_4|^2 -\frac{3}{128st^3} |f_2|^2 u -\frac{6}{128st^3} Re(f_2 f^{\star}_4)+\frac{\mu^4_1}{t^3}+\frac{\mu^4_2}{st^2}.
\end{equation}
From $\frac{\partial}{\partial u} V^{(2)}_{dS}=0$, we obtain:
\begin{equation} \label{VdS2paru}
u^2=\frac{|f_4|^2}{|f_2|^2} \quad \text{and} \quad u=\frac{|f_4|}{|f_2|}.
\end{equation}
From $\frac{\partial}{\partial s} V^{(2)}_{dS}=0$, we derive:
\begin{equation} \label{VdS2pars}
3|f_2||f_4|+3Re(f_2 f^{\star}_4)-64\mu^4_2 t=0.
\end{equation}
From $\frac{\partial}{\partial t} V^{(2)}_{dS}=0$, we get:
\begin{equation} \label{VdS2part}
9|f_2||f_4|+9Re(f_2 f^{\star}_4)-192\mu^4_1 s-128\mu^4_2 t=0.
\end{equation}
By combining \eqref{VdS2pars} and \eqref{VdS2part}, we obtain:
\begin{equation} \label{VdS4st}
3\mu^4_1s=\mu^4_2 t.
\end{equation}
Thus, the total potential becomes
\begin{equation} \label{VdS2min}
V^{(2)}_{dS}= -\frac{6 \mu^{12}_2}{128\times 27 \mu^{12}_1 s^4}Re(f_2f^{\star}_4) +\frac{4\mu^{12}_2}{27 \mu^{8}_1 s^3} - \frac{6 \mu^{12}_2}{128\times 27 \mu^{12}_1 s^4} |f_2||f_4|.
\end{equation}
If we set $V^{(2)}_{dS}=0$ and avoid fine-tuning, (i.e., $\mu^4_1=\mu^4_2=0$), we find that $s\rightarrow +\infty$, which violates the swampland distance conjecture \cite{palti2019swampland,van2022lectures,agmon2022lectures}. If we require $V^{(2)}_{dS}=0$ and $\mu^4_1=\mu^4_2=0$, the minimum total potential is  
\begin{equation} \label{VdS2min1}
V^{(2)}_{dS}= -\frac{3}{64 s t^3}Re(f_2f^{\star}_4) - \frac{3}{64 st^3} |f_2||f_4|<0.
\end{equation}
Thus, we cannot choose this particular form of $\Delta W$.

\subsubsection{$\Delta W=f_4 U + f_2 U^2 + f_0 U^3$}
Now consider the following form of $\Delta W$:
\begin{equation} \label{DeltaW3}
\Delta W^{(3)}= f_4 U + f_2 U^2 + f_0 U^3,
\end{equation}
which leads to 
\begin{equation} \label{DeltaW32}
|\Delta W^{(3)}|^2= |f_4|^2 u^2 + |f_2|^2 u^4 + |f_0|^2 u^6 + 2Re(f_2 f^{\star}_4)u^3 + 2Re(f_0 f^{\star}_4)u^4 + 2Re(f_0 f^{\star}_2)u^5
\end{equation}
and
\begin{eqnarray} \label{VAdS3}
\begin{split} 
V^{(3)}_{AdS}=  & -\frac{3}{128st^3u}|f_4|^2 -\frac{3}{128st^3} |f_2|^2 u-\frac{3}{128st^3} |f_0|^2 u^3\\
& -\frac{6}{128st^3} Re(f_2 f^{\star}_4) -\frac{6}{128st^3} Re(f_0 f^{\star}_4)u  -\frac{6}{128st^3} Re(f_0 f^{\star}_2) u^2.
\end{split}
\end{eqnarray}
Due to the complexity of the tadpole conditions \eqref{tcc1}-\eqref{tcc4}, we instead focus on the general uplift term \eqref{VbarD6}, such that the total potential is given by $V^{(3)}_{dS}=V^{(3)}_{AdS}+V_{up}$. 

From the condition $\frac{\partial}{\partial u} V^{(3)}_{dS}=0$, we obtain:
\begin{equation} \label{VdS3paru}
3|f_0|^2u^4+4Re(f_0f^{\star}_2)u^3+[2Re(f_0f^{\star}_4)+|f_2|^2]u^2-|f_4|^2=0.
\end{equation}
In the regime where $u$ is small, the terms $u^3$ and $u^4$ can be neglected. This approximation yields: 
\begin{equation} \label{VdS3paru2}
u \approx \frac{|f_4|}{\sqrt{2Re(f_0f^{\star}_4)+|f_4|^2}}.
\end{equation}
If we require $V^{(3)}_{dS} \rightarrow 0$ without fine-tuning, we again find $s \rightarrow +\infty$, similar to the case of $\Delta W_2= f_4 U + f_2 U^2$. Furthermore, if we impose $V^{(3)}_{dS}=0$ and take $\mu^4_1=\mu^4_2=0$, then due to the suppression of the term $u^3$, the minimum of the potential remains negative -- as shown earlier in \eqref{VdS2min1}. Thus, this form of $\Delta W$ is not viable.

\subsubsection{$\Delta W= f_2 U^2 + f_0 U^3$}
We now consider the following form of $\Delta W$ 
\begin{equation} \label{DeltaW}
\Delta W^{(4)}= f_2 U^2 + f_0 U^3.
\end{equation}
This leads to
\begin{equation} \label{DeltaW2}
|\Delta W^{(4)}|^2= |f_2|^2 u^4+ 2Re(f_0 f^{\star}_2) u^5 + |f_0|^2 u^6
\end{equation}
and the corresponding AdS potential becomes
\begin{equation} \label{VAdS2}
V^{(4)}_{AdS}=-\frac{3}{128st^3}|f_2|^2 u-\frac{6}{128st^3}Re(f_0 f^{\star}_2) u^2-\frac{3}{128st^3}|f_0|^2 u^3.
\end{equation}
The tadpole conditions \eqref{tcc1}-\eqref{tcc4} are difficult to apply directly in this context. Therefore, we focus on the general uplift term given in \eqref{VbarD6}. The total potential then takes the form:
\begin{eqnarray} \label{VAdS4}
\begin{split} 
V^{(4)}_{dS}=  & -\frac{3}{128st^3}|f_2|^2 u-\frac{6}{128st^3}Re(f_0 f^{\star}_2) u^2-\frac{3}{128st^3}|f_0|^2 u^3\\
& +\frac{\mu^4_1}{t^3}+\frac{\mu^4_2}{st^2}.
\end{split}
\end{eqnarray}
By solving $\frac{\partial}{\partial u} V^{(4)}_{dS}=0$, we obtain:
\begin{equation} \label{VdS4paru}
u=\frac{-2Re(f_0 f^{\star}_2) \pm \sqrt{4Re(f_0 f^{\star}_2)^2-3|f_0|^2 |f_2|^2}}{|f_0|^2}.
\end{equation}
Since $u \geq 0$, the discriminant must be non-negative, which leads to the condition:
\begin{equation} \label{ugeq0}
3|f_0|^2 |f_2|^2=0,
\end{equation}
implying that either $f_0=0$ or $f_2=0$, and consequently $u=0$. However, because $f_0$ appears in the denominator of \eqref{VdS4paru}, we must have $f_2=0$ to avoid divergence. Therefore, the expression for $\Delta W= f_2 U^2 + f_0 U^3$ reduces to $\Delta W=f_0 U^3$.

\subsubsection{$\Delta W=f_0 U^3$}
By setting $q_i=0$ (i.e., $q_1=q_2=q_3=0$), the tadpole conditions \eqref{tcc1}-\eqref{tcc4} reduce to the following expressions:
\begin{equation} \label{tcc11}
\sum_a N_a n^1_a n^2_a n^3_3 = 16-\frac{1}{2} mh_0,
\end{equation}
\begin{equation} \label{tcc21}
\sum_a N_a n^1_a m^2_a m^3_a = -16-\frac{1}{2} mh_1,
\end{equation}
\begin{equation} \label{tcc31}
\sum_a N_a m^1_a n^2_a m^3_a = -16-\frac{1}{2} mh_2,
\end{equation}
\begin{equation} \label{tcc41}
\sum_a N_a m^1_a m^2_a n^3_a = -16-\frac{1}{2} mh_3.
\end{equation}
Under these conditions, the $V_{D6/O6}$ becomes:
\begin{equation} \label{Vup5}
V^{(5)}_{D6/O6}= \frac{1}{8t^3} + \frac{3}{8st^2} - \frac{1}{256t^3}mh_0+ \frac{1}{256st^2}m\tilde{h}=V^{(5)}_{up1},
\end{equation}
where $\tilde{h}=h_1+h_2+h_3$. From the extremization condition $\frac{\partial}{\partial t} V^{(5)}_{up}=0$, we find:
\begin{equation} \label{Vup5part}
s=\frac{2m\tilde{h}+192}{3mh_0-96}t.
\end{equation}
Similarly, the condition $\frac{\partial}{\partial s} V^{(5)}_{up}=0$ gives:
\begin{equation} \label{Vup5pars}
m\tilde{h}=-96.
\end{equation}
From $\frac{\partial}{\partial t} V^{(5)}_{up}=0$, we have
\begin{equation} \label{Vup5part}
mh_0=32.
\end{equation}
Using these results, we find that the uplift potential vanishes: $V^{(5)}_{D6/O6}=0$. This implies that an additional negative contribution (a downshift) is required in order to achieve a de Sitter (or a Minkowski) minimum.

The AdS potential is given by
\begin{equation} \label{VAdS5}
V^{(5)}_{AdS}=-\frac{3}{128st^3}|f_0|^2 u^3,
\end{equation}
and the total potential is expressed as
\begin{equation} \label{VdS5}
V^{(5)}_{dS}=-\frac{3}{128st^3}|f_0|^2 u^3 +\frac{1}{8t^3} + \frac{3}{8st^2} - \frac{1}{256t^3}mh_0+ \frac{1}{256st^2}m\tilde{h}.
\end{equation}
By taking the derivative of the total potential with respect to $u$, we obtain:
\begin{equation} \label{VdS5paru1}
\frac{\partial}{\partial u} V^{(5)}_{dS}=-\frac{9}{128st^3}|f_0|^2 u^2=0.
\end{equation}
If we do not require $s$ or $t$ to tend to infinity, we conclude that $u=0$. 

Starting from the condition $\frac{\partial}{\partial t} V^{(5)}_{dS}=0$, we get the equation:
\begin{equation} \label{VdS5part1}
\frac{9}{128st^4}|f_0|^2 u^3-\frac{3}{8t^4}-\frac{6}{8st^3}+\frac{3}{256t^4}mh_0-\frac{2}{256st^3}m\tilde{h}=0.
\end{equation}
Since $u=0$, the equation simplifies to: 
\begin{equation} \label{VdS5t}
s=\frac{2m\tilde{h}+192}{3mh_0-96}t.
\end{equation}

Next, from $\frac{\partial}{\partial s} V^{(5)}_{dS}=0$, we obtain the following equation:
\begin{equation} \label{VdS5pars1}
\frac{3}{128s^2t^3}|f_0|^2 u^3-\frac{3}{8s^2t^2}-\frac{1}{256s^2t^2}m\tilde{h}=0.
\end{equation}
Again, with $u=0$, this simplifies to:
\begin{equation} \label{VdS5s}
m\tilde{h}=-96.
\end{equation}
By using equation \eqref{Vup5part}, we conclude that the minimum value of $V^{(5)}_{dS}$ is zero. 


We now extend this result to the general uplift term given in \eqref{VbarD6}. Following a similar procedure, we again find that $u=0$, $\mu^4_1=\mu^4_2=0$, which is consistent with the conditions derived from \eqref{Vup5pars} and \eqref{Vup5part}. As a result, the total potential reaches a minimum value of zero. This outcome aligns with our initial assumption that the underlying spacetime is nearly Minkowski. In this framework, both de Sitter and anti-de Sitter vacua are understood as small perturbations around a Minkowski background.

\subsubsection{$\Delta W=f_2 U^2$}
The corresponding AdS potential is given by
\begin{equation} \label{VAdS6}
V^{(6)}_{AdS}=-\frac{3}{128st^3}|f_2|^2u.
\end{equation}
In this case, applying the tadpole conditions \eqref{tcc1}-\eqref{tcc4} is not straightforward, so we focus instead on the general uplift term from \eqref{VbarD6}. The total scalar potential takes the form
\begin{equation} \label{VdS6}
V^{(6)}_{dS}=-\frac{3}{128st^3}|f_2|^2u +\frac{\mu^4_1}{t^3}+ \frac{\mu^4_2}{st^2}.
\end{equation}
Taking the derivative with respect to $u$, we find
\begin{equation} \label{VdS6paru}
\frac{\partial}{\partial u} V^{(6)}_{dS}=-\frac{3}{128st^3}|f_2|^2=0.
\end{equation}
This leads to two possible scenarios:\\
1. Either $s$ or $t$ must go to infinity, which we want to avoid;\\
2. $|f_2|^2=0$, implying $f_2=0$, which in the notation of \eqref{Wf} corresponds to $m=0$, and hence $\Delta W=0$.

Neither of these options is acceptable. Therefore, we conclude that the choice $\Delta W=f_2 U^2$ is not viable. In conclusion, to remain consistent with the Swampland Distance Conjecture, our attention should be directed toward the following scenario: 
\begin{equation} \label{DeltaW6}
\Delta W= f_0 U^3.
\end{equation}
According to \cite{kallosh2020mass}, because $\Delta W= f_0 U^3$ is relatively small, it does not affect the result of moduli stabilization. 

\subsection{Hierarchy problem}
According to equation \eqref{m232}, for the downshift scenario with $\Delta W = f_0 U^3$, the gravitino mass squared is given by
\begin{equation} \label{m232'}
m^2_{3/2}=e^K (\Delta W)^2= \frac{1}{128st^3u^3}|f_0|^2 u^6=\frac{1}{128st^3}|f_0|^2 u^3.
\end{equation}
Thus, the gravitino mass becomes
\begin{equation} \label{m32'}
m_{3/2}=\frac{1}{2^{7/2}s^{1/2}t^{3/2}}|f_0| u^{3/2},
\end{equation}
where we have set $M_{pl}=1$. As discussed in \cite{Conlon:2008zz,linde2012supersymmetry}, if the gravitino mass is below 100 TeV, supersymmetry can help address the hierarchy problem without requiring fine-tuning. We will now determine the values of $s$, $t$ and $u$ based on equation \eqref{m32'} to ensure this condition is satisfied. 

The values of $s$, $t$ and $u$ in equation \eqref{m32'} must satisfy the vacuum conditions outlined in \eqref{moduliMin}. In the scenario we are considering, the superpotential for the SUSY Minkowski vacuum takes the form
\begin{equation} \label{WSTU}
W= h_T T + r_T UT + h_S S + r_S US + f_4 U + W_{np},
\end{equation}
where $W_{np}$ is the non-perturbative contribution given in \eqref{Wi}. According to the vacuum conditions in \eqref{moduliMin}, the equations must satisfy:
\begin{eqnarray} \label{Weq0}
\begin{split} 
W=  & h_T T + r_T UT + h_S S + r_S US + f_4 U + A_T e^{-a_T T}- B_T e^{-b_T T}  \\
& + A_S e^{-a_S S}- B_S e^{-b_S S} + A_U e^{-a_U U}- B_U e^{-b_U U}=0,
\end{split}
\end{eqnarray}
\begin{equation} \label{WparT}
\partial_T W= h_T + r_T U -A_T a_T e^{-a_T T}+ B_T b_T e^{-b_T T}=0,
\end{equation}
\begin{equation} \label{WparS}
\partial_S W= h_S + r_S U -A_S a_S e^{-a_S S}+ B_S b_S e^{-b_S S}=0,
\end{equation}
\begin{equation} \label{WparU}
\partial_U W= r_T T +r_S S +f_4-A_U a_U e^{-a_U U}+ B_U b_U e^{-b_U U}=0,
\end{equation}
where for simplicity, we assume that all parameters are independent of the moduli $T$, $S$ and $U$. In fact, these parameters generally depend on the moduli \cite{cribiori2019uplifting,danielsson2014alternative}. By simultaneously solving the four equations \eqref{Weq0}-\eqref{WparU}, we can determine the vacuum expectation values of the three moduli in the SUSY Minkowski background, i.e., $t=t_0$, $s=s_0$ and $u=u_0$. Since $\Delta W$ is much smaller than $W$, we can still reliably use the values $t=t_0$, $s=s_0$ and $u=u_0$ to evaluate the gravitino mass as given in equation \eqref{m32'} \cite{kallosh2020mass}. In this work, we do not attempt to explicitly determine the parameter values; this remains an open direction for future investigation. As previously mentioned, if the gravitino mass lies below 100 TeV, supersymmetry may still provide a natural solution to the hierarchy problem without the need for fine-tuning.

\subsection{A simple example}
To illustrate the viability of our mechanism, we consider a simple toy model. We take $s \sim t \sim 100$ and $u \sim 0.1$ (since we have proved that $u$ should be small), then according to \eqref{m32'}, we have obtain 
\begin{equation} \label{m32'value}
m_{3/2} \sim |f_0| \cdot 10^{-7} M_{pl}.
\end{equation}
If we require $m_{3/2} \leq 100\text{TeV} \approx 10^{-13}M_{pl}$, we have $|f_0| \leq 10^{-6}$. Such small values can be achieved in flux compactification by discrete parameter selection, such as \cite{dewolfe2005type}. We should emphasize that the values of $t$, $s$, $u$ and other parameters are not arbitrary and should satisfy moduli stabilization conditions \eqref{Weq0}-\eqref{WparU}.

Of course, according to our above analysis, the perturbative superpotential $\Delta W=f_0 U^3$ satisfies the tadpole cancellation conditions \eqref{tcc1}-\eqref{tcc4} or \eqref{tcc11}-\eqref{tcc41} (we set $q_1=q_2=q_3=0$). More concrete D6-brane set-up can be found in \cite{cvetivc2001three} but we do not discuss in this paper. Although the Standard Model is not yet embedded, this simple example supports the viability of our construction as a first step toward realistic model building.

\section{Conclusion and Outlook}
In this paper, we have proposed a unified string-theoretic framework in which three major fine-tuning problems -- the strong CP problem, the hierarchy problem, and the cosmological constant problem -- can be simultaneously addressed. We worked within type IIA string theory compactified on a $T^6/(\mathbb{Z}_2 \times \mathbb{Z}_2)$ orientifold, incorporating RR and NS fluxes, and adopted a Kallosh-Linde-type structure for moduli stabilization.

The strong CP problem is resolved using a 4-form -- coupled axion mechanism, embedded naturally into the type IIA compactification. We showed that a specific superpotential correction of the form $\Delta W = f_0 U^3$ allows for a small and stable uplift of the vacuum energy while preserving compatibility with the Swampland Distance Conjecture. We further demonstrated that this structure leads to a naturally small gravitino mass -- potentially explaining the electroweak-Planck scale hierarchy -- without fine-tuning.

One limitation of this project that must be acknowledged is the uncertainty in the calculation of the uplifting mechanism. However, because the perturbation $\Delta W$ is very small, the resulting AdS vacua are only mildly warped rather than deep AdS vacua. This makes the uplifting process more manageable and theoretically controllable in this context compared to scenarios involving deep AdS minima. In this sense, the smallness of the cosmological constant might be seen as indirectly arising from this feature -- nature may be presenting us with a less complex problem. Only near Minkowski vacua can the calculations become easier and reliable. In this work, we have only obtained Minkowski vacua, focusing solely on the contributions from moduli fields. It is possible that including the effects of fermions and the Higgs particle could generate a small positive cosmological constant. We leave a detailed investigation of this possibility for future work.

Another reason for choosing the type IIA $T^6/(\mathbb{Z}_2 \times \mathbb{Z}_2)$ model as the starting point of this study is that, in the presence of intersecting D6-branes, this string theory setup can relatively easily realize the Standard Model of particle physics. For example, in \cite{cvetivc2001three} and subsequent papers, the authors constructed a supersymmetric version of the Standard Model that includes three generations of fermions. In addition to the particles found in the Minimal Supersymmetric Standard Model (MSSM), the model also accommodates right-handed neutrinos and other additional particles. 

While our current model is not yet a full realization of the Standard Model in string theory, it provides a consistent proof-of-principle that these longstanding problems may share a common stringy origin. A simple example with concrete flux values and moduli vevs confirms the viability of the construction. In future work, we aim to:
\begin{itemize}
  \item Embed realistic chiral matter content, such as MSSM-like spectra, using intersecting D-brane configurations;
  \item Study the specific mechanism of strong CP problem in type IIA $T^6/(\mathbb{Z}_2 \times \mathbb{Z}_2)$ model since this paper has not yet studied the specific mechanism; 
  \item Investigate the structure of Yukawa couplings and fermion mass hierarchies within this compactification;
  \item Study the dynamics of axion dark matter and the cosmological evolution of moduli in this framework;
  \item Explore implications for low-scale supersymmetry breaking and possible experimental signatures.
\end{itemize}
We believe that this framework provides a promising starting point for constructing phenomenologically viable models grounded in string theory.

\acknowledgments

YL was supported by an STFC studentship. Thanks for the discussion with Antonio Padilla, Paul Saffin and Benjamin Muntz. For the purpose of open access, the authors have applied a CC BY public copyright licence to any Author Accepted Manuscript version arising.




\begin{thebibliography}{99}

\bibitem{bousso2000quantization}
R. Bousso and J. Polchinski, \emph{JHEP} {06} (2000)006.

\bibitem{kachru2003sitter}
S. Kachru, R. Kallosh, A.D. Linde and S.P. Trivedi, \emph{Phys. Rev. D} {68} (2003)046005.

\bibitem{giddings2002hierarchies}
S.B. Giddings, S. Kachru and J. Polchinski, \emph{Phys. Rev. D} {66} (2002)106006.

\bibitem{dvali2006large}
G. Dvali, \emph{Phys. Rev. D} {74} (2006)025018.

\bibitem{dvali2005three}
G. Dvali, hep-th/0507215. 

\bibitem{dvali2006vacuum}
G. Dvali, \emph{Phys. Rev. D} {74} (2006)025019.

\bibitem{dvali2014neutrino}
G. Dvali, S. Folkerts and A. Franca,\emph{Phys. Rev. D} {89} (2014)105025.

\bibitem{dvali2022strong}
Gia Dvali, arXiv:2209.14219, 2022. 

\bibitem{burgess2024uv}
C.P.Burgess, Gongjun Choi and F. Quevedo, \emph{JHEP} {03} (2024)051.

\bibitem{choi2023implications}
Gongjun Choi and Jacob Leedom, \emph{JHEP} {09} (2023)175.

\bibitem{kachru2003sitter}
S. Kachru, R. Kallosh, A.D. Linde and S.P. Trivedi, \emph{Phys. Rev. D} {68} (2003)046005. 

\bibitem{balasubramanian2005systematics}
V. Balasubramanian, P. Berglund, J.P. Conlon and F. Quevedo, \emph{JHEP} {03} (2005)007.

\bibitem{conlon2005large}
J.P. Conlon, F. Quevedo and K. Suruliz, \emph{JHEP} {08} (2005)007.

\bibitem{Liu:2023vqp}
Yang Liu, Antonio Padilla and Francisco G. Pedro, \emph{JHEP} {10} (2023)014.

\bibitem{Liu:2024blx}
Yang Liu, Antonio Padilla and Francisco G. Pedro, \emph{JHEP} {08} (2024)048.

\bibitem{grimm2005effective}
T.W. Grimm and J. Louis, \emph{Nucl. Phys. B} {718} (2005)153.

\bibitem{louis2002type}
J. Louis and A. Micu, \emph{Nucl. Phys. B} {635} (2002)395.

\bibitem{villadoro2005N}
G. Villadoro and F. Zwirner, \emph{JHEP} {06} (2005)047.

\bibitem{dewolfe2005type}
O. DeWolfe, A. Giryavets, S. Kachru and W. Taylor, \emph{JHEP} {07} (2005)066.

\bibitem{camara2005fluxes}
P.G. Camara, A. Font and L.E. Ibanez, \emph{JHEP} {09} (2005)013.

\bibitem{ibanez2012string}
Ibanez, Luis E and Uranga, Angel M, String theory and particle physics: An introduction to string phenomenology, Cambridge University Press, 2012.

\bibitem{bielleman2015minkowski}
Sjoerd Bielleman, Luis E. Ibanez and Irene Valenzuela, \emph{JHEP} {12} (2015)119.

\bibitem{bergshoeff2001new}
E. Bergshoeff, R. Kallosh, T. Ortin, D. Roest and A. Van Proeyen, \emph{Class. Quant. Grav.} {18} (2001)3359.

\bibitem{cribiori2019uplifting}
Niccolo Cribiori, Renata Kallosh, Christoph Roupec and Timm Wrase, \emph{JHEP} {12} (2019)171.

\bibitem{kallosh2020mass}
Renata Kallosh and Andrei Linde, \emph{JHEP} {01} (2020)169.

\bibitem{dibitetto2011charting}
G. Dibitetto, A. Guarino and D. Roest, \emph{JHEP} {03} (2011)137.

\bibitem{danielsson2014alternative}
U. Danielsson and G. Dibitetto, \emph{JHEP} {05} (2014)013.



\bibitem{Conlon:2008zz}
Joseph P. Conlon, \emph{Mod.Phys.Lett.A} {23} (2008)1-16.

\bibitem{linde2012supersymmetry}
Andrei Linde, Yann Mambrini, and Keith A. Olive, \emph{Phys. Rev. D} {85} (2012) 066005.

\bibitem{kutasov2015constraining}
David Kutasov, Travis Maxfield, Ilarion Melnikov, and Savdeep Sethi, \emph{Phys.Rev.Lett.} {115} (2015) 7, 071305.

\bibitem{cvetivc2001three}
Mirjam Cvetic, Gary Shiu and Angel M. Uranga, \emph{Phys.Rev.Lett.} {87} (2001) 20, 201801.

\bibitem{conlon2006qcd}
Joseph P. Conlon, \emph{JHEP} {05} (2006) 078. 

\bibitem{palti2019swampland}
Eran Palti, \emph{Fortschritte der Physik} 67(6), (2019):1900037.

\bibitem{van2022lectures}
Marieke van Beest, José Calderón-Infante, Delaram Mirfendereski and Irene Valenzuela, \emph{Physics Report} 989 (2022) 1-50.

\bibitem{agmon2022lectures}
Nathan Benjamin Agmon, Alek Bedroya, Monica Jinwoo Kang, and Cumrun Vafa, arXiv:2212.06187, 2022. 


\end{thebibliography}


\end{document}